\documentclass[a4paper]{jpconf}
\usepackage{graphicx}

\usepackage{graphics,graphicx}
\usepackage{multicol} 
\usepackage{parskip}
\usepackage{amsmath}
\usepackage{multirow}
\usepackage{caption}
\usepackage{subcaption}

\begin{document}

\begin{center}
    {\fontsize{14bp}{20bp}\selectfont\it Contribution to the 25th International Workshop on Neutrinos from Accelerators}
\end{center}

\vspace{-27mm} 

\title{ProtoDUNE Photon Detection System}

\author{J. Soto-Oton on behalf of the DUNE Collaboration}

\address{IFIC-CSIC/UV, Carrer del Catedrátic José Beltrán Martinez, 2, 46980 Paterna, Spain}

\ead{j.soto@cern.ch}

\begin{abstract}
The Deep Underground Neutrino Experiment (DUNE) is a long-baseline neutrino oscillation experiment aiming to measure the oscillation parameters with an unprecedented precision that will allow determining the CP violation phase in the leptonic sector and the neutrino mass ordering. The Far Detector of DUNE will consist of four 17 kton liquid argon Time Projection Chambers (LAr-TPC). Inside a LAr-TPC, a Photon Detection System (PDS) is needed to detect the scintillation light produced by the interacting particles. The PDS signal provides the interaction time for non-beam events and improves the calorimetric reconstruction. To validate DUNE technology, two large-scale prototypes, of 750 ton of LAr each, have been constructed at CERN, ProtoDUNE-HD and ProtoDUNE-VD. The PDS of both prototypes is based on the XArapuca concept, a SiPM-based device that provides good detection efficiency covering large surfaces at a reasonable cost. This document presents the preliminary performance of the ProtoDUNE-HD Photon Detection System, which has taken data from April to November 2024.
\end{abstract}

\section{Introduction}

DUNE\cite{DUNEtdrv1} will be a long-baseline neutrino oscillation experiment that will perform precision measurements of the PMNS mixing parameters, unambiguously determine the mass order, and discover leptonic CP violation\cite{DUNEOscillations}. It also comprises a rich non-accelerator physics program such as the detection of supernova\cite{DUNESN}, atmospheric and solar neutrinos, as well as BSM physics searches. 
DUNE will consist of the most powerful neutrino beam, produced at Fermilab (US), a Near Detector placed downstream of the beam, and a Far Detector located 1.5 km underground at the Sanford Underground Research Facility (SURF), around 1,300 km away from the beam source. The Far Detector will consist of four LAr-TPCs  of 17\,kton with dimensions of 12\,m\,x\,12\,m\,x\,60\,m each. 

The LAr-TPC detector technology enables 3D particle imaging with millimetric resolution. In a LAr-TPC, the interacting particles ionize the liquid argon along their tracks. Then, the ionization electrons are drifted by a uniform electric field and read out by a segmented anode. The excellent imaging capabilities of the LAr-TPC technology, and the advantages of using liquid argon as a detector medium for neutrino physics, will allow DUNE to fulfil its physics goals. Additionally, scintillation light is produced in liquid argon, which is detected by a dedicated PDS (Photon Detection System), which provides a key feature of a TPC: It is used to reconstruct the particle interaction time needed to determine the drift distance inside the TPC. This is required to perform an unambiguous 3D reconstruction which allows the background rejection of particles entering the detector from the outside. Also, as it provides a fast signal, it can be used as a trigger for non-beam events. However, scintillation light in LAr is produced at a wavelength of 127 nm, in the VUV (Vacuum UltraViolet) range, at which most photo-sensors are not sensitive. To address these challenges, the DUNE collaboration developed the concept of X-Arapuca\cite{Machado:2018rfb}, which combines the use of wavelength shifters (WLS) and dichroic filters to trap photons, shift their wavelength towards the visible range, and redirect them to SiPMs (silicon photo-multipliers).

DUNE will bring the LAr-TPC technology to a massive scale. The first two LAr-TPCS of the Far Detector will have two different designs: In the Horizontal Drift (HD) approach, the active volume is segmented into four volumes with a horizontal drift field. The maximum drift length is 3.6\,m. In the VD design, the volume is divided in two, placing a cathode in the middle and drifting the electrons vertically towards the two anodes, which are placed at the top and the bottom. To validate DUNE technology, two large-scale prototypes, of 750 tons of LAr each, have been constructed at CERN: ProtoDUNE-HD (for Horizontal Drift), which took data from March 2024 to November 2024, and ProtoDUNE-VD (for Vertical Drift) which will take data during the first quarter of 2025.

\section{ProtoDUNE-HD Photon Detection System}
ProtoDUNE-HD aims to test the Far Detector components at a real scale. In ProtoDUNE-HD, four anodes are placed on the sides of the active volume, and a cathode is in the centre. The active volume is 4.6\,m\,x\,6.1\,m\,x\,7.2 m, providing a maximum horizontal drift of 3.6\,m as in the FD.
The anodes are four Anode Planes Assemblies (APAs), with 3 sensing layers of wires with a pitch of 4.7\,mm. 40 X-Arapucas are placed behind each APA to detect scintillation photons, counting 160 in total. Each X-Arapuca is 48\,cm x\,10\,cm, and contains 48 electrically-ganged SiPMs.

Four different X-Arapuca configurations are used in ProtoDUNE-HD, by combining two different WLS manufacturers (Eljen and G2P) and two different SIPM models (HPK and FBK).
Dedicated measurements have been performed at the laboratory to characterize the photon detection efficiency (PDE) of the different configurations, resulting in a slightly better performance for X-Arapucas with the combination of G2P+HPK \cite{Alvarez-Garrote:2024bvd}. Additionally, ProtoDUNE-HD SiPMs have been extensively characterized in the laboratory \cite{Andreotti:2024blh}. The PDS signal readout is performed by the DAPHNE\cite{esteban} boards, the analog-front end system that manage the digitization and triggering of the cold amplifier channels.

The filling of ProtoDUNE-HD started in March 2024, and the operations in May 2024. It was exposed to a beam of charged particles for 10 weeks from June to September 2024. The beam is composed of electrons, protons, pions, kaons and muons with momenta between 0.5 GeV/c and 7 GeV/c. ProtoDUNE-HD has collected around 30 million beam and cosmic interactions.

\subsection{SiPM calibration and monitoring}

\begin{figure}[ht]
     \centering
     \begin{subfigure}[b]{0.40\textwidth}
         \centering
         \includegraphics[width=\textwidth]{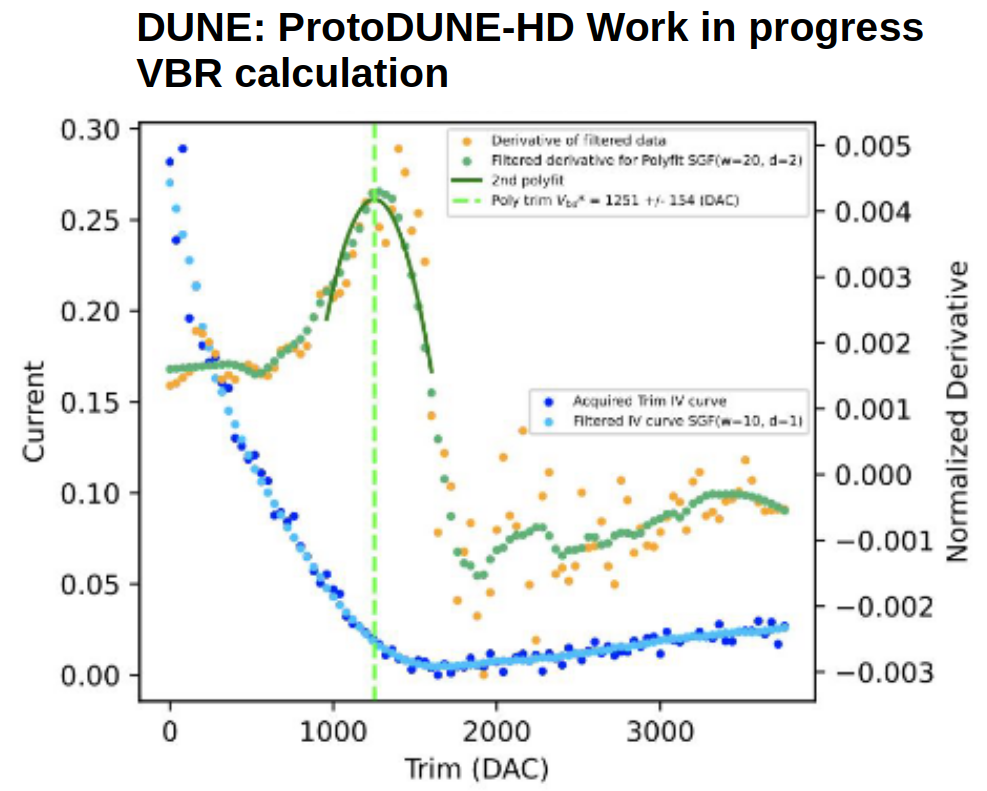}
     \end{subfigure}
     \hfill
     \begin{subfigure}[b]{0.58\textwidth}
         \centering
         \includegraphics[width=\textwidth]{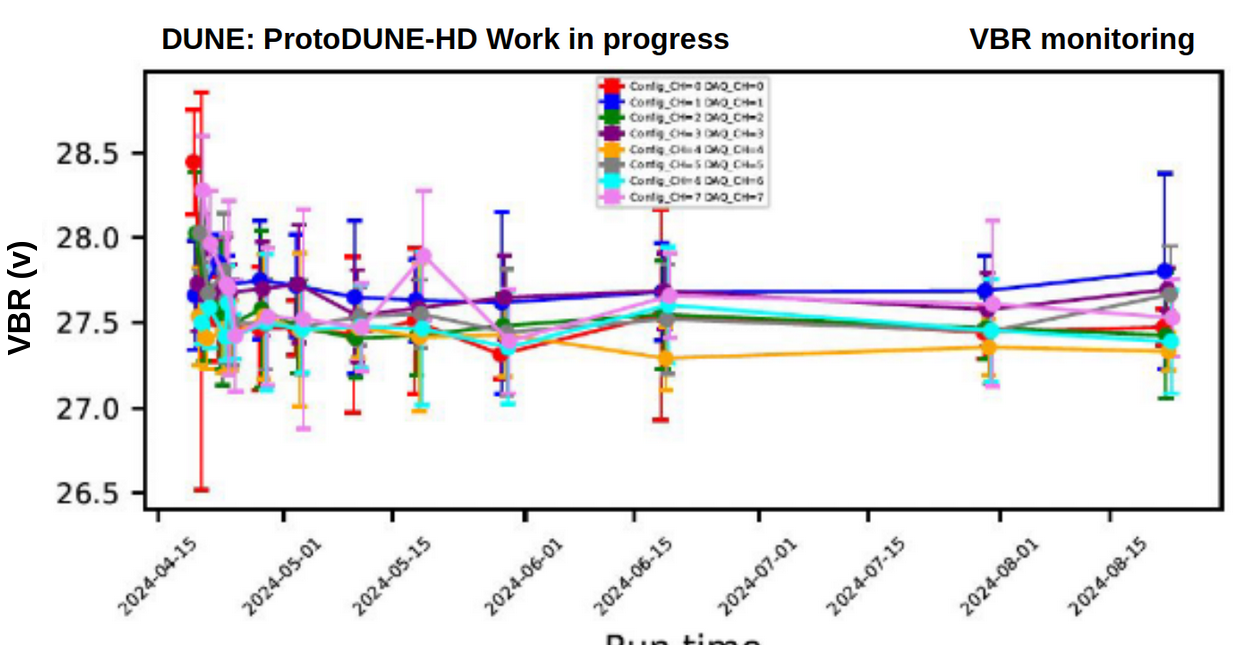}
     \end{subfigure}
        \caption{Left: IV curve from a ProtoDUNE-HD SiPM in light (dark) blue for raw (filtered) data. The logarithmic derivative is shown in orange for the filtered data, and in green the filtered derivative. In dark green is shown the polynomial fit around the local maximum. Right: Evolution of VBR for a selection of SiPMs.}
        \label{fig:VBR}
\end{figure}

The DAPHNE boards allow performing dedicated current-voltage (IV) scans to compute and monitor the SiPM break-down voltage (VBR). Figure \ref{fig:VBR} left shows how the VBR is obtained by computing the logarithmic derivative of the current. Then a second-order polynomial fit is performed around the local maximum to obtain the VBR value. The VBR is measured weekly and the operation voltage is adjusted to ensure a uniform detection efficiency across all channels. Figure \ref{fig:VBR} shows the stability of the VBR from April until August 2024.


A dedicated LED-based calibration system is used to periodically monitor the gain and the Signal-To-Noise ratio of all SiPMs. Additionally, the DAPHNE Front End amplifier can be tuned on a per-channel basis to equalize the response for all SiPMs.

\subsection{Tau slow monitoring}
\begin{figure}[ht]
     \centering
     \begin{subfigure}[b]{0.40\textwidth}
         \centering
         \includegraphics[width=\textwidth]{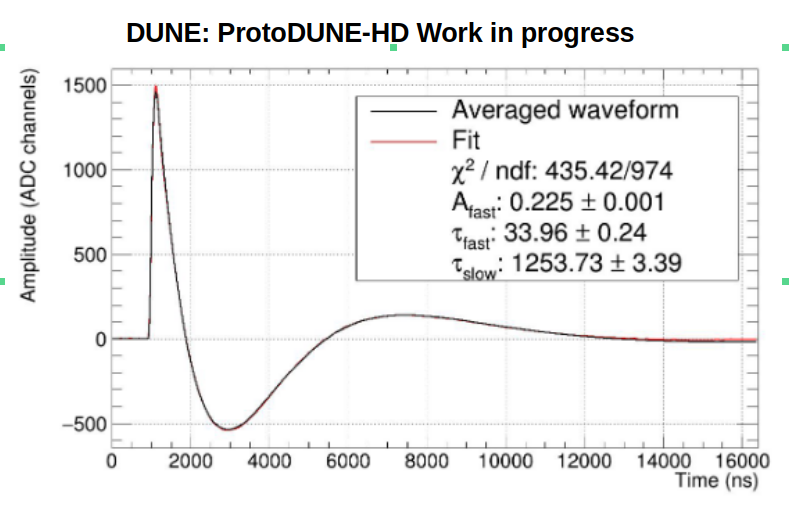}
     \end{subfigure}
     \hfill
     \begin{subfigure}[b]{0.58\textwidth}
         \centering
         \includegraphics[width=\textwidth]{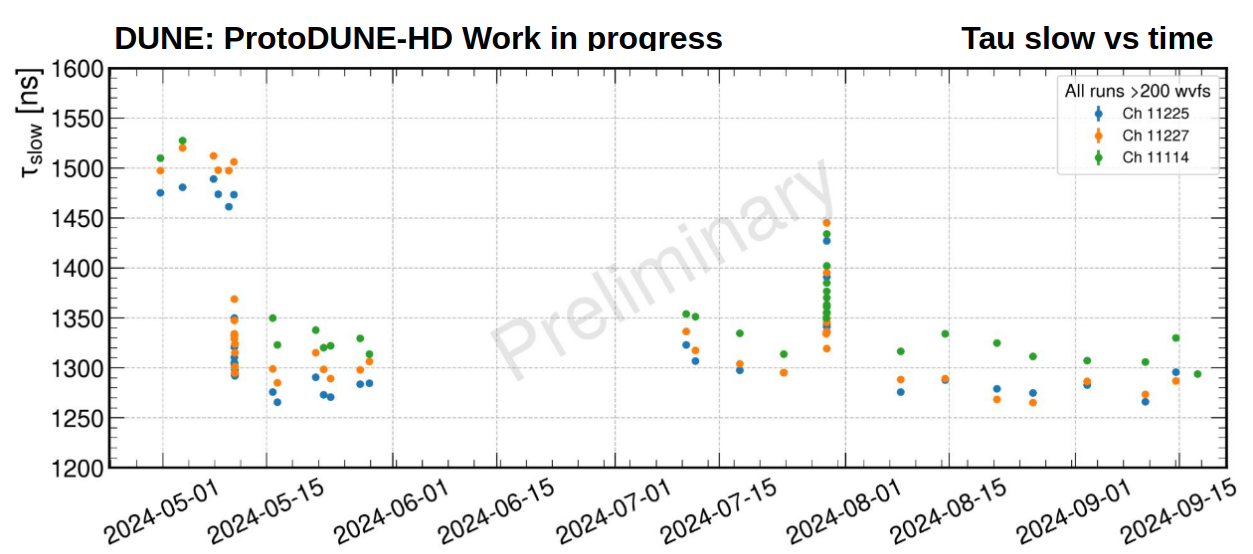}
     \end{subfigure}
        \caption{Left: Example of an average waveform fit to two exponential functions convoluted with the SPE response. Right: Evolution of $\tau_{slow}$ for a selection of channels. The data until the second week of May is shown at 0\,kV/cm of drift field, then the nominal drift field of 0.5\,kV/cm is shown. The spike at the end of July corresponds to a drift-field scan.}
        \label{fig:tau}
\end{figure}

Scintillation light in pure liquid argon is produced by the radiative decay of molecular argon excimers to the ground state. This radiative decay occurs with two characteristic times, a fast decay of $\tau_{\text{fast}}$= 7 ns and a slow decay of around $\tau_{\text{slow}}$ =1.5 $\mu$s. The value of $\tau_{\text{slow}}$ provides an indication of the liquid argon purity, since contaminants in liquid argon, such as nitrogen or oxygen, lead to the reduction of the scintillation light production due to a quenching process driven by the two-body collision of excited argon excimers with the impurities. The value of  $\tau_{\text{slow}}$ is extracted from the data by fitting the sum of two exponential functions convoluted with the SiPM response. The left panel of Figure \ref{fig:tau} shows an example of the fit, with a resulting value of around 1.3 $\mu$s under the presence of the drift field. The right panel of Figure \ref{fig:tau} shows the evolution of $\tau_{slow}$ measured during the detector operation for a selection of three channels. The data shown at the beginning of the plot is taken without the drift field and shows an average value of around 1.5 $\mu s$. This value indicates a high purity, corresponding to an equivalent of $0.48\pm0.04$\,ppm of nitrogen, in agreement with the expectations. In the second week of May 2024, the drift field was activated, and the average value decreased to around 1.3 $\mu$s. This decrease with the drift field is in agreement with previous measurements \cite{DUNE:2022ctp,Aimard:2020qqa} reported in the literature.

\subsection{Light yield vs beam energy}
\begin{figure}[ht]
     \centering
     \begin{subfigure}[b]{0.59\textwidth}
         \centering
         \includegraphics[width=\textwidth]{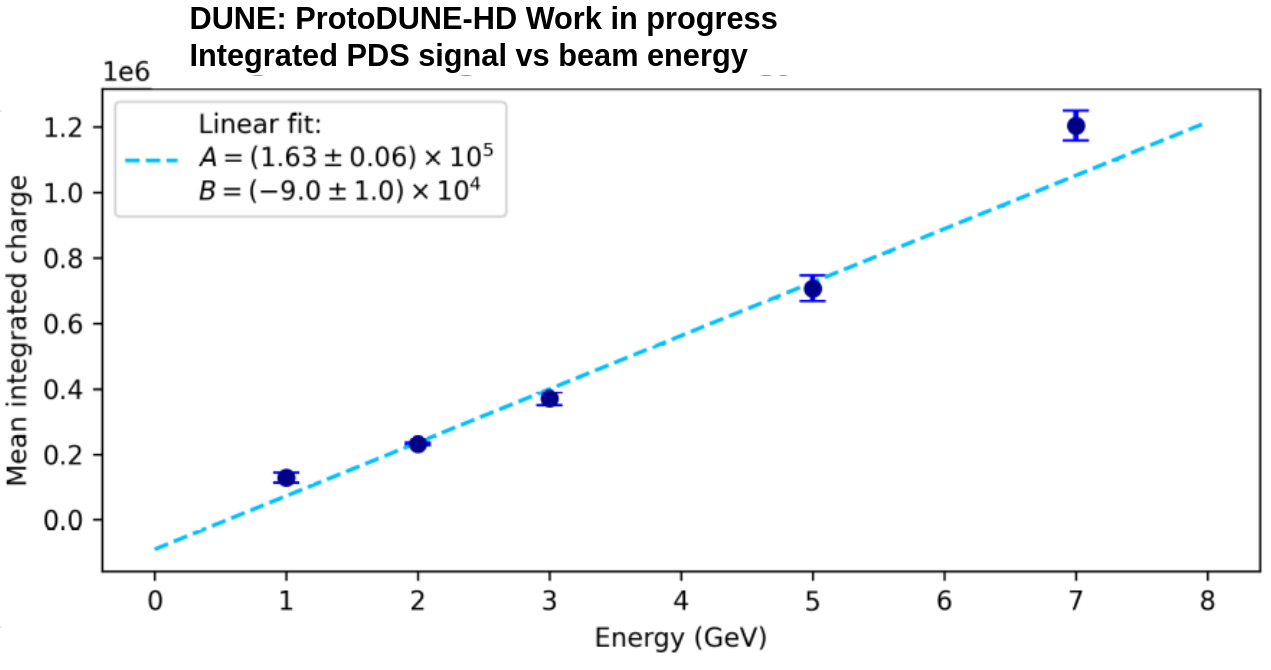}
     \end{subfigure}
     \hfill
     \begin{subfigure}[b]{0.39\textwidth}
         \centering
         \includegraphics[width=\textwidth]{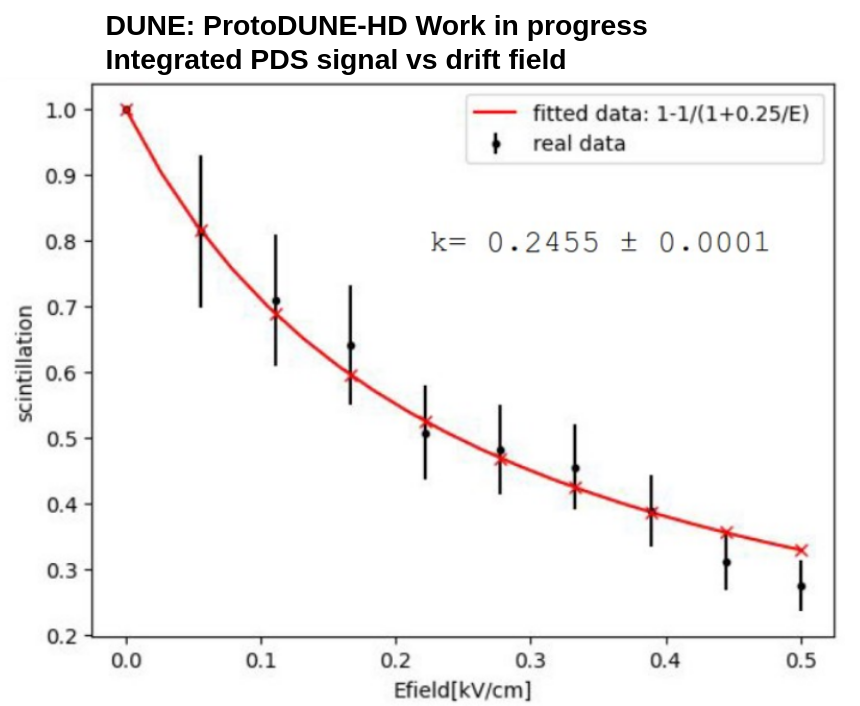}
     \end{subfigure}
        \caption{Left: Dependence of the detected light with the beam energy. Right: Dependence of the detected light with the drift field.}
        \label{fig:beam}
\end{figure}

Since the amount of scintillation light produced is proportional to the deposited energy, the PDS can provide an independent estimate of the particle energy, contributing to the calorimetric reconstruction. The left panel of Figure \ref{fig:beam} shows the preliminary results of the dependence of the integrated PDS signal with the beam energy, showing good linearity. At this stage, no cuts have been applied.

\subsection{Drift field studies}
The dependence of the photon production on the drift field is also being studied. In the absence of a drift field, the ionization electrons recombine producing excimers that decay producing more scintillation light. Therefore, a decrease in light production with the drift field is expected. The right panel of Figure \ref{fig:beam} shows the dependence of the detected light with the drift field. A reduction is observed in agreement with the expectations.

\section{Conclusions}
ProtoDUNE-HD has been taking beam and cosmic data from April to November 2024. The ProtoDUNE-HD PDS has been stably operated during all the data-taking. The preliminary results of the measurement of the $\tau_{\text{slow}}$ and on the linearity of the detected light with the particle energy are in agreement with the expectations.

\section*{Acknowledgments}
The author acknowledges support from the Generalitat Valenciana of Spain under grant PROMETEO/2021/087.
\section*{References}
\bibliographystyle{iopart-num}
\bibliography{biblio}

\end{document}